\documentclass[a4paper]{article}
\usepackage{ASVspoofVoicePrivacy}
\usepackage{subfig}
\usepackage{graphicx,epstopdf}
\usepackage{epsfig,amssymb,amsmath}
\usepackage{xcolor}
\ninept
\PassOptionsToPackage{hyphens}{url}
\usepackage[bookmarks=false,hidelinks]{hyperref}
\usepackage{pgfplots}
\DeclareUnicodeCharacter{2212}{−}
\usepgfplotslibrary{groupplots,dateplot}
\usetikzlibrary{patterns,shapes.arrows}
\pgfplotsset{compat=newest,scaled y ticks=false}

\title{Benchmarking and challenges in security and privacy for voice biometrics}

\makeatletter
\def\name#1{\gdef\@name{#1\\}}
\makeatother
\name{{\em Jean-Francois Bonastre, Héctor Delgado, Nicholas Evans, Tomi Kinnunen,}\\
      {\em Kong Aik Lee, Xuechen Liu, Andreas Nautsch, Paul-Gauthier Noé, Jose Patino,}\\
      {\em Md Sahidullah, Brij Mohan Lal Srivastava, Massimiliano Todisco,}\\ 
      {\em Natalia Tomashenko, Emmanuel Vincent, Xin Wang, Junichi Yamagishi}}

\address{ASVspoof and VoicePrivacy organising committees \\
{\small \tt organisers@asvspoof.org, organisers@voiceprivacychallenge.org} }
\begin{document}
\maketitle

\begin{abstract}

For many decades, research in speech technologies has focused upon improving reliability.
With this now meeting user expectations for a range of diverse applications, speech technology is today omni-present.
As result, a focus on security and privacy has now come to the fore.
Here, the research effort is in its relative infancy and progress calls for greater, multi-disciplinary collaboration with security, privacy, legal and ethical experts among others.
Such collaboration is now underway.
To help catalyse the efforts, this paper provides a high-level overview of some related research. 
It targets the non-speech audience and  
describes the benchmarking methodology that has spearheaded progress in traditional research and which now drives recent security and privacy initiatives related to voice biometrics.
We describe: the ASVspoof challenge relating to the development of spoofing countermeasures; the VoicePrivacy initiative which promotes research in anonymisation for privacy preservation.

\end{abstract}

\section{Introduction}

The voice is among the most natural and convenient means to human-machine interaction and biometric authentication.  In some scenarios, particularly telephony or teleconferencing applications, voice can be the \emph{only} available biometric. While automatic speaker verification (ASV) systems can provide a reliable means to authentication, like all biometric technologies, it is not without security and privacy concerns.  Arguably, these concerns are potentially greater for voice biometrics than they can be for authentication systems that use alternative biometric characteristics.

Security concerns relate to the potential for ASV systems to be manipulated by adversaries through spoofing attacks~\cite{sahidullah2019introduction}, now referred to as presentation attacks~\cite{ISOpresentationAtack}.  Fraudsters can launch spoofing attacks to gain illegitimate access to protected services or resources by presenting to the ASV system a speech recording which has been manipulated to \emph{sound}\footnote{We refer to machine perception rather than human perception; humans and machines do not \emph{hear} in the same way.} like another speaker.  Without adequate protection, spoofing attacks can substantially degrade the reliability of almost any ASV system.

While not related specifically to voice biometrics, but to speech technology more generally, privacy concerns relate to the potential for speech data to be exploited for purposes other than those to which an individual might have given consent~\cite{NautschGDPR}.  Speech signals are a rich source of personal, private information.  In providing recordings of speech to a particular voice service, the speaker usually furnishes the service provider with much more information than is strictly necessary in order to perform the expected task, hence the need for privacy preservation.

In this article we describe two specific benchmarking challenges launched by the speech processing research community to expedite solutions to security and privacy concerns.  The first involves solutions to protect ASV systems from being manipulated by spoofing in the form of spoofing countermeasures or presentation attack detection systems~\cite{ISOpresentationAtack} whose development is spearheaded through the ASVspoof initiative launched in 2015~\cite{wu2015asvspoof, Kinnunen2017, asvspoof2019}.  The second relates to the VoicePrivacy initiative, launched in 2020~\cite{tomashenko2020introducing}, which aims to promote the development of privacy preservation solutions for speech technology.  The inaugural VoicePrivacy challenge focused upon anonymisation, namely techniques to manipulate speech data in order that it cannot be used with ASV systems to recognise the speaker.

The article is intended as a contribution to the efforts to build bridges between the speech community and, e.g., the legal and ethical communities and, in particular, to support the recently formed Security and Privacy in Speech Communication (SPSC) special interest group of the International Speech Communication Association (ISCA). It targets the non-specialist and is hence intentionally high-level, with a focus upon the essentials and clarity, rather than upon scientific and technical rigour.  We describe the importance of benchmarking campaigns, classifier fundamentals and the early, classical approaches to performance estimation.  After presenting a high-level overview of ASV systems, the remainder of the article introduces the ASVspoof and VoicePrivacy initiatives.

\section{Evaluation-driven research}

In the early days, researchers collected their own datasets to develop methods (algorithms and software). To make technological progress in complex tasks such as speech or speaker recognition, and to be able to meaningfully compare different methods, the need for commensurable performance benchmarking was quickly recognised. From the mid-1990s, the National Institute of Standards and Technology (NIST) --- a US-based standardisation body --- pioneered \emph{evaluation-driven} research~\cite{lee2020a,Greenberg2020}. The key ingredients are: (1)~commonly agreed (often public) data, evaluation rules, and performance metrics; (2)~disentangled roles for researchers (evaluees) and evaluators. Performance claims should not be reported by evaluees but instead by independent evaluators who provide common infrastructure (data, rules, metrics) in the form of an \emph{evaluation campaign} or \emph{challenge}. Only with such a level playing field can competing methods can be meaningfully compared. There are many such challenges~\cite{Nagrani2020,Zeinali2020} within the speech field. Typically, they require participants~\cite{i4u2019,JHUMIT2019} to design special-purpose software to solve some specific tasks.  For many, the software takes the form of a \emph{binary classifier}, namely a system which must choose between two mutually exclusive hypotheses.

Even if classifiers are based on well-established statistical principles and are trained objectively via numerical optimisation techniques, they can (and do) make errors. 
Errors are the result of over-simplified modelling assumptions, natural variability in the input data, sources of external nuisance variation, or as a result of limited training data, etc. The design of ever-more reliable classifiers is the traditional bread and butter of machine learning and speech research.

A binary classifier makes two types of errors: \emph{misses} (false rejections) and \emph{false alarms} (false acceptances). Misses imply that the classifier rejects a positive input that should be accepted. A false alarm results from the classifier accepting a negative input that should be rejected. Classifier performance can be estimated by dividing the number of misses and false alarms by the number of tested positive and negative cases respectively. The resulting \emph{miss rate} ($P_{miss}$) and \emph{false alarm rate} ($P_{FA}$) can be seen as proxies for user convenience and security.  

\begin{figure}[t]
    \centering
    \subfloat[]{\includegraphics[trim={5cm 5.5cm 6cm 4.5cm},clip,width=0.9\linewidth]{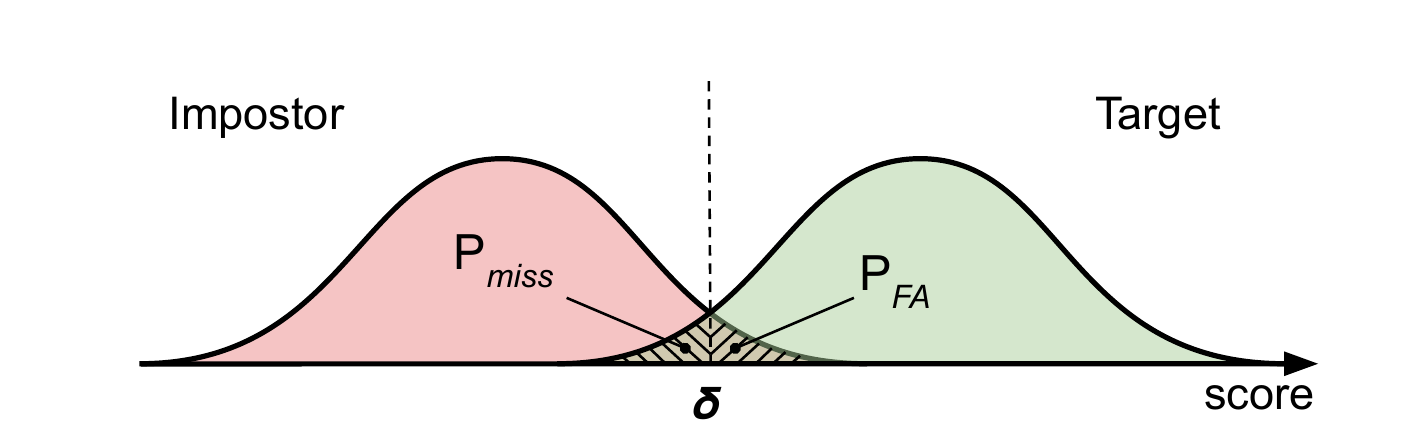}}\\
    \subfloat[]{\includegraphics[trim={5cm 5cm 6cm 4.5cm},clip,width=0.9\linewidth]{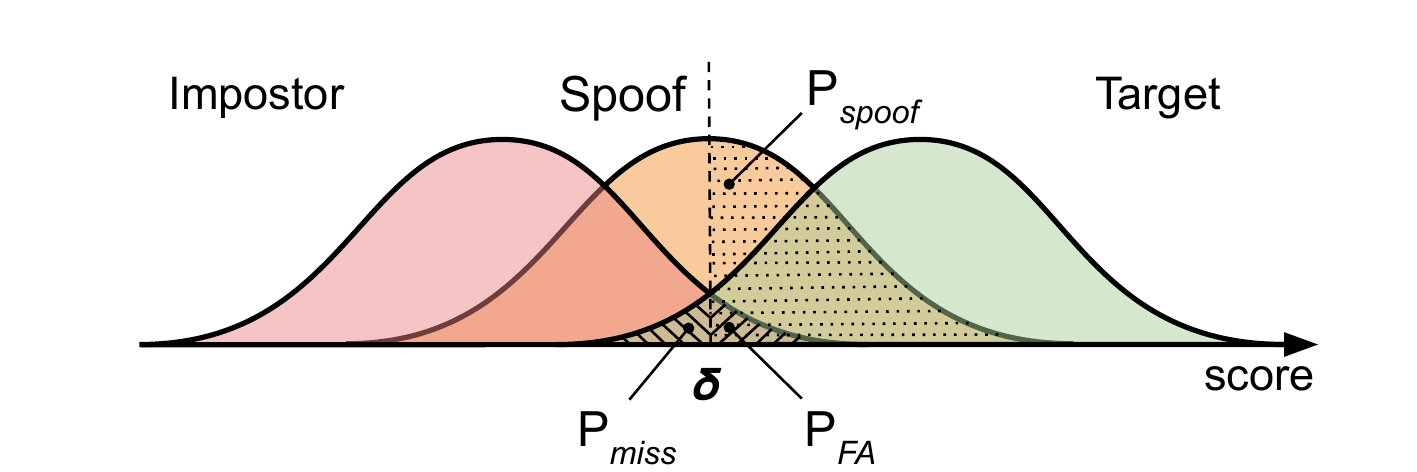}}\\
    \subfloat[]{\includegraphics[trim={5cm 5.5cm 6cm 4.5cm},clip,width=0.9\linewidth]{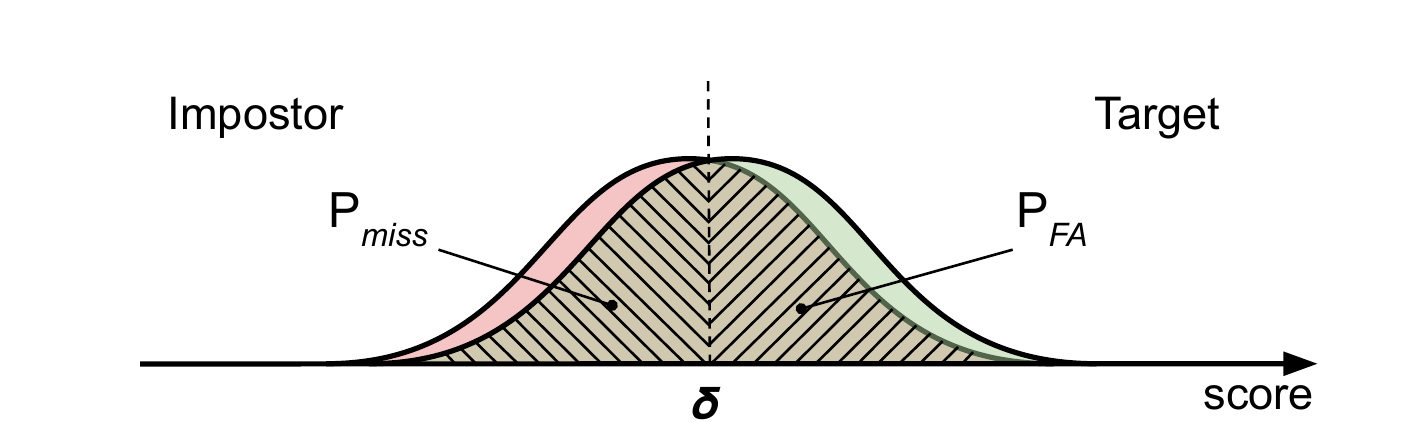}}
    \caption{Illustrative score distributions for (a)~automatic speaker verification (ASV), (b)~an ASV system subjected to spoofing attacks and (c)~an ASV system presented with anonymised data.} 
    \label{fig:distributions}
\end{figure}

Binary classifier decisions are derived in two stages. The first involves computation of a `soft decision' (the \emph{score}) --- a real number that expresses the classifier's confidence in the positive case.  Example raw score \emph{distributions} for such a binary classifier are illustrated in Fig.~\ref{fig:distributions}(a) were the positive case is the target class and the negative case is the impostor class. In practice, the score is often a \emph{logarithmic likelihood ratio} (LLR) and expresses the relative strength between the two competing hypotheses. Second, the score is compared with a pre-set \emph{threshold} $\delta$. Scores above the threshold imply `accept', whereas scores below the threshold imply `reject'. By increasing the threshold, the false alarm rate ($P_{FA}$, red shaded area in Fig.~\ref{fig:distributions}) can be reduced at the expense of a higher miss rate ($P_{miss}$, green shaded area) and vice versa.  The trade-off can be readily visualised in so-called detection error trade-off (DET) plots~\cite{martin1997det} such as that illustrated in Fig.~\ref{fig:asvspoof}~(described below).

Since there are two error rates, which do we report?  Or do we report both?  What threshold / how do we set it?  The answer to these questions is rather subtle and a detailed treatment is outside the scope of this paper. One particular metric, namely the \emph{equal error rate}~(EER) is adopted for a broad range of tasks.  It is computed using the threshold that makes miss and false alarm rates equal (see Fig.~\ref{fig:asvspoof}) --- thereby yielding a single number to report. The \emph{lower} the EER, the more reliable the classifier.

The authors acknowledge that, even if the 
EER is deprecated in ISO standards~\cite{ISO-IEC-197951-IS-2021}, it provides a compact summary of the \emph{discrimination} capabilities of a classifier --- how well it is capable of observing (in speech, `hearing') differences between positive and negative inputs.  In practice, however, the EER does not provide a full picture.  More comprehensive approaches to assessment have been developed and have been broadly adopted by the community.  These provide a view of classifier performance through the lens of a formal \emph{decision policy} which represents the effective trade-off between the two decision outcomes~\cite{andreas2019speaker}.

\section{Automatic speaker verification}
\label{sec:asv}

ASV systems provide one of the most natural and convenient means to biometric person authentication. Test recordings (probes) are compared with enrolment recordings (references) to verify (or not) a claimed identity. The reference is used to create a model\footnote{
    On account of their dynamic nature and the variability in speech signals, we refer to models, not templates.
} 
which is stored in a reference database. At test time, the model corresponding to the claimed identity is compared to the test utterance 
resulting in a soft score.  A hard accept/reject decision can be fully automated (e.g., online banking) or semi-automated (with some human intervention, e.g., when forensic practitioners present voice evidence in court). Scores should reflect reliably the extent to which the strength-of-evidence supports the same/different identity propositions: the higher the score, the greater the similarity and vice versa.

Detection error trade-off plots for two different ASV systems assessed using the ASVspoof 2019 logical access database and the VoicePrivacy 2020 database are illustrated in Figs.~\ref{fig:asvspoof} and~\ref{fig:voiceprivacy} and show EERs of 2.5\% and 1.1\% respectively (blue profiles).  Each point on any one profile corresponds to a different decision threshold (operating point).  The profiles show how misses can be traded off against false alarms in order to meet different application requirements.

\section{Security vulnerabilities: ASVspoof}

\begin{figure}[!t]
  \centering
  \includegraphics[trim={12.5cm 0 13cm 0},clip,width=0.7\linewidth]{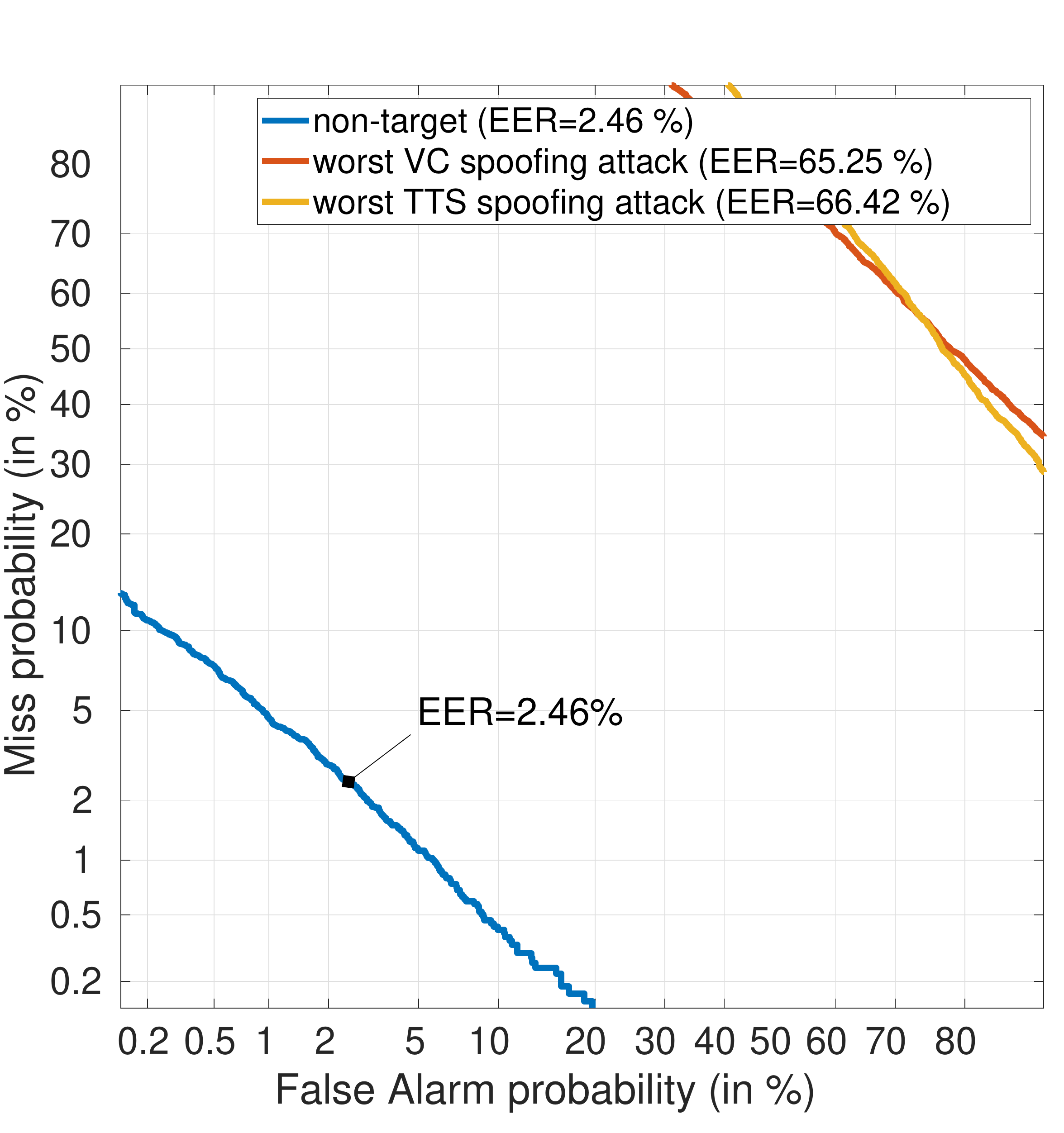}
  \caption{Detection error trade-off plot for the ASVspoof 2019 logical access task.  Profiles shown for the baseline {\bf ASV} system (blue profile) and the same system subjected to the most effective synthetic (text-to-speech, TTS) speech spoofing attack (orange profile) and the most effective voice conversion (VC) spoofing attack (red profile).} 
  \label{fig:asvspoof}
\end{figure}

Without adequate protections, the reliability of ASV systems can be compromised by the presentation of synthetic or converted voice, replayed speech and impersonation~\cite{asvspoof2019}.  Generated automatically from a text input, today's state-of-the-art synthesis systems are capable of producing speech that the human cannot distinguish from bona fide speech~\cite{shen2018natural}.  Voice conversion systems operate directly upon an input speech signal and alter the voice to that of another speaker~\cite{Yi2020}.  Unlike synthetic and converted voice spoofing attacks, which both demand a certain technical expertise and suitable training and adaptation data, replay attacks can be launched by the layman, requiring only consumer-grade recording and replaying devices.  All can substantially degrade ASV reliability.  Impersonation, while still a threat~\cite{hautamaki2013vectors, Wu2015}, is less effective.

The impact of spoofing attacks upon an ASV system is illustrated in Fig.~\ref{fig:distributions}(b).  Spoofing attacks introduce a third class of input so that the ASV system must now cope with target (matching), impostor (non-matching) and spoofed utterances.  A successful spoofing attack circumvents the ASV system by provoking a score above the decision threshold.  The score distribution for spoofing attacks is illustrated in the middle of Fig.~\ref{fig:distributions}(b).  Spoofing attacks provoke a false alarm rate $P_{spoof}$ that is greater than the original false alarm rate $P_{FA}$.  One can think of spoofing attacks as a special case of impostors where there is a concerted effort to deceive the ASV system.  

Without the capacity to distinguish between spoofed and bona fide speech, ASV reliability will degrade as a result of spoofing attacks, sometimes substantially.  This degradation assessed using the ASVspoof 2019 logical access database (synthetic and converted voice spoofing attacks) is illustrated in Fig.~\ref{fig:asvspoof}.  The baseline EER (with no spoofing attacks) of 2.5\% increases to over 50\% when impostor trials are replaced by the most effective synthetic and converted voice spoofing attacks.  These results should be interpreted with caution, however.  In practice, one must consider the relative likelihood of the ASV system being presented with target and impostor trials, versus that of spoofing attacks.  Without this consideration, the results in Fig.~\ref{fig:asvspoof} might convey an overly pessimistic view of the vulnerabilities to spoofing.  There is, in any case, potential to detect attacks automatically using countermeasures.

\begin{figure}[!t]
  \centering
  \includegraphics[trim={13cm 0 13cm 0},clip,width=0.7\linewidth]{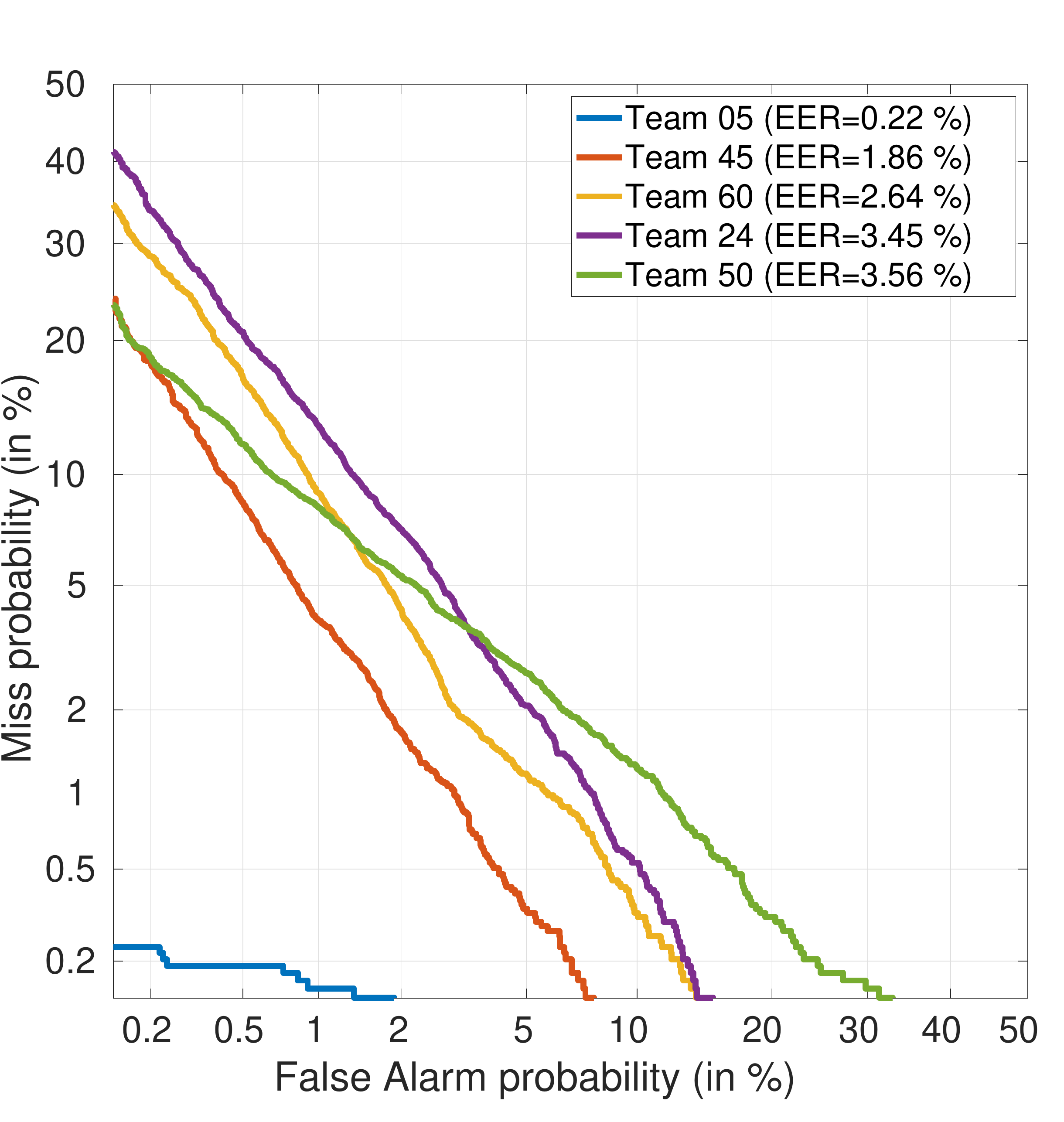}
  \caption{As for Fig.~\ref{fig:asvspoof} except for {\bf spoofing countermeasures}.  Profiles shown for the top five performing systems.}
  \label{fig:asvspoof_cm}
\end{figure}

The series of ASVspoof challenges held bi-annually since 2015 have spearheaded the development of spoofing countermeasures or presentation attack detection (PAD) solutions for ASV.  Spoofing countermeasures can be applied prior to ASV in order to detect attacks and prevent them from reaching the ASV system.  The most recently completed challenge was held in 2019 and included separate logical access (synthetic and converted voice attacks) and physical access (replay attacks) tasks~\cite{asvspoof2019}.  DET plots for the top-five performing spoofing countermeasures for the ASVspoof 2019 logical access task are shown in Fig.~\ref{fig:asvspoof_cm}.  They show EERs of as low as 0.2\%.  

Thus, while spoofing attacks can present a substantial threat to reliability, and while human listeners may not be able to detect spoofing attacks, countermeasures can be effective.  Even so, while ASV countermeasures have proven potential to detect attacks, they can also degrade usability; they can erroneously classify bona fide speech as spoofed speech.  Assessment and performance estimates should hence reflect the impact of both spoofing and countermeasures \emph{upon the ASV system}; countermeasures should be assessed in tandem with ASV.  Such more elaborate approaches to assessment, e.g.~\cite{kinnunen2020tandem}, are outside the scope of the current article.

\section{Privacy implications: VoicePrivacy}

Speech signals contain much more than just the spoken message or the voice identity~\cite{NautschCSLPrivacy,COMPRISE_D5.1}.  The speaker's sex/gender, age, socio-economic and geographical background, emotion and health condition etc.\ can all be estimated automatically using recordings of speech~\cite{shafran2003voice, schultz2007speaker}. With much of this information being personal and private there is hence an interest to develop privacy safeguards for speech technology.  This is the goal of the VoicePrivacy initiative, founded in 2020. While solutions to privacy preservation can take many different forms, and while VoicePrivacy may explore different approaches in the future, the inaugural challenge focused upon the development of \emph{anonymization} solutions~\cite{tomashenko2020introducing}.  

The idea is to protect privacy by distorting speech data such that it cannot be used by ASV systems to recognise the speaker.  Anonymisation is achieved by suppressing the information in speech signals that is typically used by machines to infer identity.  On its own, this is relatively straightforward, e.g.\ by adding sufficient levels of background noise, or by replacing speech with silence.  The challenge comes from the requirement to suppress personally identifiable attributes contained within the speech signal while leaving all other attributes intact.  These requirements imply that anonymisation should not interfere with the application of some down stream tasks such as automatic speech recognition, nor should it introduce processing artefacts that might degrade subjective intelligibility or naturalness.  Last, anonymised voices should remain distinctive, meaning the task might better be referred to as pseudonymisation.

Rather than operating upon the speech signal so that it reflects the voice of another, specific speaker, anonymisation aims to prevent the speaker identity from being recognised.  Anonymisation acts to increase the confusion between utterances produced by the same and different speakers.  For a perfect anonymisation system, the score distributions corresponding to impostor and target trials overlap.  The desired effect of anonymisation upon an ASV system is illustrated in Fig.~\ref{fig:distributions}(c).  In this case, the ASV system cannot produce simultaneously both a low false alarm rate and a low miss rate, no matter what the decision threshold, and the EER is 50\%.  Real anonymisation solutions are less effective. 

\begin{figure}[t]
  \centering
  \includegraphics[trim={13cm 0.5 13cm 0},clip,width=0.7\linewidth]{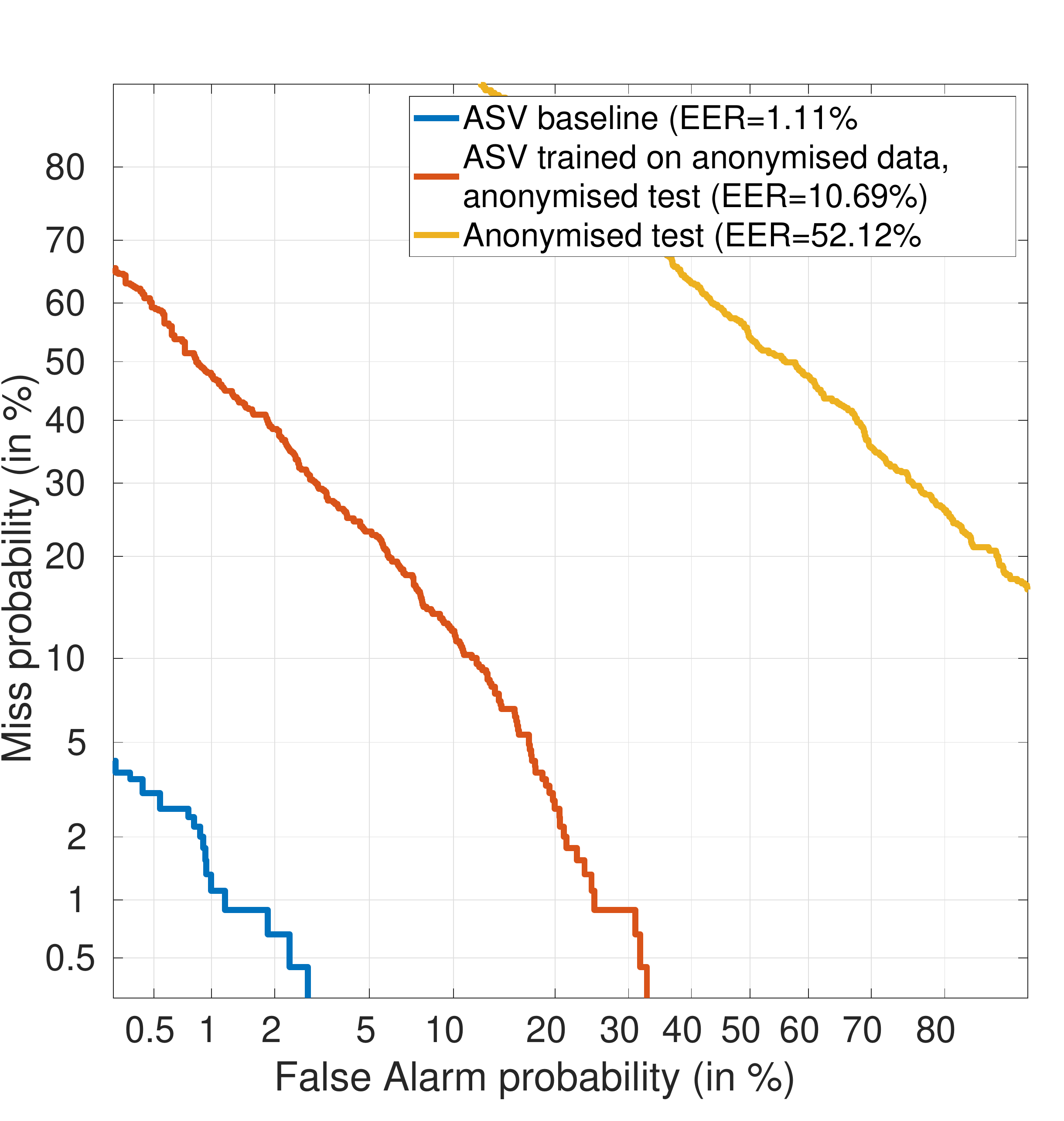}
    \caption{Detection error trade-off plot for the VoicePrivacy 2020 challenge (LibriSpeech test, male trials).  Profiles shown for the baseline ASV system (blue profile) and the same system presented with anonymised test utterances (orange profile) and anonymised test utterances when the ASV system is re-trained using similarly anonymised training data (red profile).}
  \label{fig:voiceprivacy}
\end{figure}

The first VoicePrivacy challenge was held in 2020 and attracted submissions from 7 independent teams.   A DET plot for the primary challenge baseline~\cite{tomashenko2020introducing} is illustrated in Fig.~\ref{fig:voiceprivacy}.  It shows an increase in the EER from a baseline of 1\% (blue profile) to 52\% after anonymisation (orange profile).  The EER of over 50\% suggests that the anonymisation goal is met.  However, if an anonymisation adversary were to adapt the ASV system in light of anonymisation, then performance is less effective; anonymised utterances still contain some personally identifiable information (PII) and the potential to re-identify the speaker remains.  When the ASV system is re-trained or adapted using similarly anonymised training data, the result is a lower EER of 
10.7\%
(red profile); true anonymisation remains elusive.

The above treatment does not reflect impacts upon intelligibility/naturalness, nor voice distinctiveness.  In practice, the multiple objectives and complex nature of anonymisation means several different metrics are used in practice.  The development of more suitable approaches to assessment are the focus of current research~\cite{srivastava2019privacy,Nautsch2020,Maouche2020,Noe2020}.

\section{Reflections and further considerations}

The community has made substantial progress to address the security and privacy implications of speech technology and rapid progress has been made in the last half-decade.  We have: proposed definitions for a number of distinct tasks and solutions; established evaluation driven research initiatives as a vehicle to successfully raise the profile of security and privacy research and to build new research communities; developed protocols, criteria and metrics for the assessment of security and privacy safeguards. Nonetheless, we are perhaps still far from having a comprehensive appreciation of the implications as well as the potential of safeguards.

Will spoofing ever be a solved problem?  Possibly not.  Speech synthesis and voice conversion are long-established research fields with genuine applications, such as anonymisation, yet the same technology poses a threat to the security of ASV.  The progress speech synthesis and voice conversion in recent years has been impressive.  Today's technology produces synthetic speech that humans cannot distinguish from bona fide speech.  Whether or not similar advances may one day result in machines that produce synthetic speech that other machines cannot detect is an intriguing question.  For the time being it is clear that, if we seek reliable, secure approaches to person authentication using ASV, we must intensify our efforts in anti-spoofing.  Future directions include a focus on more adversarial attacks, the development of countermeasures that function reliably in the wild, e.g.\ in the face of background noise and other sources of nuisance variation such as bandwidth and channel variability which typify telephony ASV scenarios, as well as the approach used to estimate performance.

The research effort in anonymisation is relatively embryonic.  While there is an opportunity to provide some level of protection, current solutions fall short of delivering true anonymisation.  Furthermore, we have observed differences in the level of privacy that a given anonymisation solution provides to different individuals.  Whereas the protection for some can be strong, others are left with relatively little protection at all.  Since personally identifiable information encapsulates far more than that used by most ASV systems to infer identity, since other alternative attributes can be used instead, and since current anonymisation solutions do not necessarily suppress them, is \emph{full} anonymisation even technically possible?  

Anonymisation solutions that focus only upon the attributes used by typical ASV systems already degrade intelligibility and naturalness.  If \emph{all} personally identifiable information can be successfully suppressed, then what remains?  Is personally identifiable information so inextricably, indelibly embedded in speech that it cannot be (fully) removed?  Even if a speech signal can be fully anonymised, will anything resembling speech remain?  Or, for a given application, what level of intelligibility/naturalness can be sacrificed in order to achieve anonymisation?  How should we even measure anonymisation performance and privacy?  Are the methodology and metrics adequate?  How does this research fit with broader solutions to privacy preservation, e.g.\ encryption and distributed learning?

That we have certainly raised more questions above than we have answered serves to show that our community's journey in security and privacy research is only just beginning.  Our efforts must be redoubled looking to the future.

\section{Acknowledgements}

The work reported in this paper was supported by:
the French ANR (VoicePrivacy, VoicePersonae, DEEP-PRIVACY, Harpocrates); the European Commission (H2020 COMPRISE); the Japan Science and Technology Agency (JST) with grant No. JPMJCR18A6; JSPS KAKENHI No.21K17775; the Academy of Finland (proj. 309629); Region Grand Est, France.

\bibliographystyle{IEEEbib}
\bibliography{ASVspoofVoicePrivacy}

\end{document}